\documentclass[aps,prD,twocolumn,superscriptaddress]{revtex4}
\usepackage{amssymb}
\usepackage{amsmath}
\usepackage{xcolor}

\newcommand{\e}{{\rm e}}   % 2.718281828

\newcommand{\rd}{{\rm d}}

\begin{document}

\title{Why Barriola--Vilenkin Global Monopoles Cannot Rotate?}
\author{Yi Lu}
\author{Xiao-Yin Pan}
\email{panxiaoyin@nbu.edu.cn}
\affiliation{Department of Physics, Ningbo University, Ningbo 315211, Zhejiang Province, China}

\author{Meng-Yun Lai}
\email{mengyunlai@jxnu.edu.cn}
\affiliation{College of Physics and Communication Electronics, Jiangxi Normal University, Nanchang 330022,  Jiangxi Province, China}

\author{Qing-hai Wang}
\email{qhwang@nus.edu.sg}
\affiliation{Department of Physics, National University of Singapore, Singapore 117551, Singapore}

\date{\today}

\begin{abstract} 
The Barriola--Vilenkin global monopoles are topological defects predicted by certain grand unified theories and have been extensively studied for their astrophysical and cosmological implications, including their distinctive spacetime geometry and characteristic gravitational lensing effects. Despite this interest, an exact solution for a global monopole remains elusive, with research largely confined to approximations of the static, spherically symmetric case. This paper addresses the fundamental question of whether a rotating global monopole can exist as a solution to the coupled Einstein-scalar field equations. We first prove that metrics generated by applying the Newman-Janis algorithm to the static monopole are inconsistent with the scalar field's equation of motion. Furthermore, we perform an asymptotic analysis for general static, axially symmetric spacetimes and establish that the only such solution that is regular at large distances is the spherically symmetric one. These results lead to the conclusion that the Barriola--Vilenkin global monopoles are incompatible with rotating spacetime within the framework of Einstein's general relativity.
\end{abstract}

%\pacs{03.65.Yz, 03.67.Lx, 03.67.Mn}
% \keywords{enatanglement;
%qubits}
\maketitle

\section{Introduction}

The Barriola--Vilenkin (BV) global monopoles are point-like topological defects hypothesized to have formed during phase transitions in the early universe \cite{Kibble1976,Vilenkin1985,Vilenkin1989,Vilenkin1994,Kibble1995}. The canonical model, first proposed by Barriola and Vilenkin, describes a static global monopole via a triplet of scalar fields with a global $\text{O}(3)$ symmetry spontaneously broken to $\text{U}(1)$ \cite{Vilenkin1989}. Their pioneering work demonstrated that the monopole's gravitational field is described by a static, spherically symmetric metric that imposes a solid angle deficit on spacetime. This distinctive feature, along with other unique physical properties, has made the global monopole a subject of considerable attention for its astrophysical and cosmological implications (see Refs.~\cite{Vilenkin1994,Kibble1995,Perlick2004,Tsujikawa2006,Radu2015} for reviews).

Following the study of the static case, a natural line of inquiry has been the search for a rotating BV global monopole solution. Several attempts have been made, including slowly rotating approximations \cite{Jing2013} and solutions generated via complex coordinate transformations \cite{Bezerra2001,Bezerra2012}. The solution proposed by Bezerra et al. \cite{Bezerra2001}, derived using a method analogous to the Newman-Janis algorithm (NJA), became particularly influential, inspiring subsequent investigations into its physical properties, such as gravitational lensing \cite{Jusufi2017,Asada2019,Jusufi2020}. However, the validity of this solution was challenged nearly two decades later \cite{Secuk2020}, sparking a debate centered on whether the proposed metric and scalar field configuration truly satisfy the scalar equations of motion and Einstein field equations \cite{Bezerra2020}.

The core of this controversy lies in the application of the NJA \cite{Newman1965a,Newman1965b}, a powerful technique for generating rotating solutions from static seeds in black hole physics \cite{Szekeres2000,Erbin2017}. A critical weakness in previous analyses is that the full set of Einstein equations for the generated rotating monopole spacetime was never explicitly verified. In this work, we resolve this long-standing issue. We derive the complete equations of motion for a scalar field within a general rotating metric generated by the modified NJA \cite{AA2014c}. We rigorously prove that no rotating global monopole solution can satisfy these equations. Consequently, we find that previously reported rotating solutions are invalid. To generalize this result, we then analyze the global monopole model within the most general static, axially symmetric metric framework \cite{Rezzolla2016}. An asymptotic analysis at large distances reveals that the only consistent solution is the spherically symmetric one. This provides strong evidence that rotating BV global monopoles are incompatible with Einstein's general relativity.

%Throughout the paper we use metric signature $(-,+,+,+)$ and the units $G=c=1$, where $G$ is  Newton's constant and $c$ is the speed of light.

\section{The static BV global monopole}

We begin by reviewing the static and spherically symmetric global monopole (GM) model as proposed by Barriola and Vilenkin \cite{Vilenkin1989}. The action for the coupled Einstein-scalar system is composed of the Einstein-Hilbert term and the matter term for the monopole:
\begin{align}
	S=\frac{1}{16\pi G}\int \rd^4x \sqrt{-g}R -\int \rd^4x \sqrt{-g} \mathcal{L}_\text{GM},
	\label{eqn:action}
\end{align}
where $g_{\mu\nu}$ is the metric tensor, $g$ is its determinant, $R$ is the Ricci scalar. The Lagrangian density for the GM, $\mathcal{L}_\text{GM}$, describes a triplet of scalar fields $\chi^i$ with a spontaneously broken global $\text{O}(3)$ symmetry:
\begin{align}
 	\mathcal{L}_\text{GM}=\frac{1}{2} (\partial_\mu \chi^i) (\partial^\mu \chi^i)+\frac{\lambda}{4}(\chi^i\chi^i-\eta^2)^2.
 	\label{eqn:LGM}
\end{align}
Here, $\lambda$ is the self-coupling constant and $\eta$ is the scale of gauge symmetry breaking. For a static monopole, the scalar fields adopt the `hedgehog' ansatz,
\begin{align}
	\begin{split}
		\chi^1 &= \eta h(r) \sin\theta \cos \phi, \\
		\chi^2 &= \eta h(r) \sin\theta \sin \phi, \\  
		\chi^3 &= \eta h(r) \cos\theta,
	\end{split}
	\label{eqn:chi}
\end{align}
where the profile function $h(r)$ approaches the vacuum manifold at spatial infinity $h(\infty) \to 1$.

Varying the action (\ref{eqn:action}) with respect to the scalar fields and the metric yields the equations of motion (EoM),
\begin{align}
	g^{\mu\nu}\nabla_\mu \nabla_\nu\chi^i =\lambda \chi^i\left(\chi^j\chi^j-\eta^2\right),\quad  i,j=1,2,3,
	\label{eqn:EoM}
\end{align}
and the Einstein field equations,
\begin{align}
	G_{\mu\nu} = 8\pi T_{\mu\nu},
	\label{eqn:FieldEquation}
\end{align}
where the energy-momentum tensor for the GM field is given by
\begin{align}
	\begin{split}
	 	T_{\mu\nu} ={}&(\partial_\mu \chi^i) (\partial_\nu \chi^i)\\
	 	& - g_{\mu\nu}\left[\frac{1}{2} (\partial_\alpha \chi^i) (\partial^\alpha \chi^i)+\frac{\lambda}{4}(\chi^i\chi^i-\eta^2)^2\right].
	\end{split}
\end{align}

To investigate rotating solutions, we adopt a stationary and axisymmetric spacetime described in Boyer-Lindquist-like coordinates $(t, r, \theta, \phi)$. The metric components $g_{\mu\nu}$ are functions of $r$ and $\theta$ only, with the only non-zero off-diagonal component being $g_{t\phi}$.  In this framework, the system is described by six non-trivial components of the Einstein field equations $(tt, t\phi, rr, r\theta, \theta\theta, \phi\phi)$. Furthermore, the scalar field equations of motion for $\chi^1$ and $\chi^2$ become degenerate, yielding one unique equation, while the EoM for $\chi^3$ provides a second. In total, this constitutes a formidable system of eight coupled, non-linear partial differential equations that must be solved simultaneously.

\section{Analysis of NJA-Generated Metrics}

We investigate the existence of a rotating global monopole within the framework of the most general stationary, axisymmetric metric in Boyer-Lindquist-like coordinates generated via the modified Newman-Janis algorithm (NJA). As derived in Ref.~\cite{AA2014c}, this metric is given by:
\begin{align}
	\begin{split}
		\rd s^2 ={}&\frac{\Psi(r,\theta)}{\rho^2} \left\{-\frac{\Delta}{\rho^2} \left(\rd t-a \sin^2\theta \rd \phi\right)^2 + \frac{\rho^2}{\Delta} \rd r^2 \right.\\
		& \left. +\rho^2 \rd \theta^2 +\frac{\sin^2\theta}{\rho^2} \left[a\rd t-\left(K+a^2\right) \rd \phi\right]^2\right\},	
	\end{split}
	\label{eqn:metric}
\end{align}
where the metric functions are defined as
\begin{align}
	\begin{split}
		K(r) &\triangleq H(r) \sqrt{\frac{F(r)}{G(r)}}, \\
		\Delta (r) &\triangleq a^2 + F(r) H(r), \\
		\rho^2(r,\theta) &\triangleq K(r) + a^2 \cos^2\theta.
	\end{split}
\end{align}
For this metric, the $r\theta$ component of the Einstein tensor, which is crucial for our analysis, is 
\begin{align}
	G_{r\theta} = 3a^2 \sin\theta \cos\theta \frac{\partial_r \rho^2}{\rho^4} - \frac{\partial_r \partial_\theta \Psi}{\Psi} + \frac{3}{2} \frac{\partial_r \Psi}{\Psi} \frac{\partial_\theta \Psi}{\Psi}.
	\label{eqn:Grtheta}
\end{align}

Our first step is to analyze the scalar field's equations of motion (EoMs). Substituting the hedgehog ansatz (\ref{eqn:chi}) into the general EoM (\ref{eqn:EoM}) yields two distinct differential equations for the profile function $h(r)$. The EoMs for the $\chi^1$ and $\chi^2$ components are degenerate and result in: 
\begin{align}
	\begin{split}
		&\Delta h''(r) + \left[(FH)' - \Delta \left(\frac{\partial_r\rho^2}{\rho^2} - \frac{\partial_r \Psi}{\Psi}\right) \right] h'(r) \\
		& + \left(\frac{a^2}{\Delta} - 2 \frac{K}{\rho^2} + \cot\theta\frac{\partial_\theta \Psi}{\Psi}\right) h(r) \\
		={}&\lambda \eta^2 \Psi h(r) \left[h^2(r)-1\right]. 
	\end{split}	
	\label{eqn:EoMxy}
\end{align}
However, due to the asymmetric form of the components in Eq.~(\ref{eqn:chi}), the EoM for $\chi^3$ is slightly different:
\begin{align}
	\begin{split}
		& \Delta h''(r) +\left[(FH)' - \Delta \left(\frac{\partial_r\rho^2}{\rho^2} - \frac{\partial_r \Psi}{\Psi}\right) \right] h'(r)\\
		&  - \left( 2\frac{a^2 + K}{\rho^2} + \tan\theta\frac{\partial_\theta \Psi}{\Psi} \right)h(r) \\
		={}& \lambda \eta^2 \Psi h(r) \left[h^2(r)-1\right]. 
	\end{split}	
	\label{eqn:EoMz}
\end{align}
For a consistent solution to exist, these two equations must be identical. Equating the coefficients of the $h(r)$ terms imposes a strict constraint on the metric function $\Psi(r,\theta)$: 
\begin{align}
	2a^2 \sin\theta\cos\theta \left(\frac{1}{2\Delta} + \frac{1}{\rho^2}\right) + \frac{\partial_\theta \Psi}{\Psi}=0.
\end{align}
This first-order linear differential equation for $\Psi$ is readily solved, yielding:
\begin{align}
	\Psi(r,\theta) = \rho^2 \e^{\frac{a^2\cos^2\theta}{2\Delta} + J(r)},
	\label{eqn:Psi}
\end{align}
where $J(r)$ is an arbitrary integrating function of $r$.

Next, we impose a constraint from the Einstein equations. A key feature of the hedgehog ansatz in this metric is that the $r\theta$ component of the energy-momentum tensor vanishes,  $T_{r\theta} = 0$. Consequently, the corresponding Einstein equation reduces to a vacuum equation,  $G_{r\theta} = 0$. Substituting our solution for $\Psi$ from Eq.~(\ref{eqn:Psi}) into the expression for $G_{r\theta}$ in Eq.~(\ref{eqn:Grtheta}) yields a differential equation for $J(r)$: 
\begin{align}
	\begin{split}
		J'(r)  ={}& \frac{\Delta'}{\Delta} \left(\frac{a^2 \cos^2\theta}{2\Delta} -2 \right) - \frac{\partial_r\left(\rho^2-4\Delta\right)}{\rho^2 + 2 \Delta } .
	\end{split}
\end{align}
A fundamental inconsistency arises: the left-hand side depends only on $r$, while the right-hand side explicitly depends on $\theta$. For a solution to exist, the right-hand side must be independent of 
$\theta$. This is trivially satisfied for the non-rotating case $(a=0)$. For a rotating solution $(a\neq0)$, all $\theta$-dependent terms must systematically cancel. By expanding the right-hand side in powers of $\cos^2\theta$ yields
\begin{align}
	\begin{split}
		J'(r) ={}&-2\frac{\Delta'}{\Delta} + \frac{\Delta'}{2\Delta^2} a^2 \cos^2\theta\\
		&- \frac{\left(K-4FH\right)'}{K + 2FH + 2a^2} \sum_{n=0}^\infty \left(-\frac{ a^2 \cos^2\theta}{K + 2FH + 2a^2}\right)^n.
	\end{split}
\end{align}  
Demand that the coefficient of each power vanishes, we find this requires: 
\begin{align}
	\Delta'=0, \qquad (K-4FH)'=0. 
\end{align}
The only non-trivial solutions to these conditions are 
\begin{align}
	F(r) = \frac{B}{H(r)}, \quad G(r) = A H(r), \quad J(r) = \ln C,
\end{align}
where $A,B,C>0$ are three integration constants. 

Finally, we substitute these constraints back into the now-consistent scalar field EoM (\ref{eqn:EoMz}) to check for final consistency. This yields the ultimate equation for the profile function $h(r)$: 
\begin{align}
	\begin{split}
		& \Delta h''(r) + \left(-2 +\frac{a^2}{\Delta} - \frac{a^2\cos^2\theta}{\Delta} \right) h(r)\\ 
		%		 =\,& \lambda \eta^2 h(r) \left[h^2(r) -1 \right] C\e^{\frac{a^2\cos^2\theta}{2\Delta}}\\
		={}& \lambda \eta^2 h(r) \left[h^2(r) -1 \right] C \sum_{n=0}^\infty \frac{1}{n!} \left(\frac{a^2\cos^2\theta}{2\Delta}\right)^n.
	\end{split}
	\label{eqn:hr}
\end{align}
This equation presents the final, insurmountable contradiction. The left-hand side contains terms linear in $\cos^2\theta$, while the right-hand side contains an infinite series in powers of $\cos^2\theta$. These distinct functional dependencies on $\theta$ cannot be reconciled for any non-zero rotation $(a\neq0)$ unless $h(r)=0$, which corresponds to the trivial case of no monopole. We note that the vacuum solutions $h(r)=\pm1$, sometimes assumed in the literature, also fail to satisfy Eq.~(\ref{eqn:hr}). Furthermore, we can see that the slow-rotation solution such that in Ref.~\cite{Jing2013} was obtaining by ignoring high-order terms in $a^2$ on the right-hand side of Eq.~(\ref{eqn:hr}).

We are therefore forced to conclude that no rotating global monopole solution can be constructed within this general class of NJA-generated metrics. This result underscores the necessity of rigorously verifying all solutions generated by the NJA against the full set of field equations.

\section{Analysis in a General Axially Symmetric Spacetime}

The analysis in the previous section, while conclusive for its class, was restricted to metrics generated by the NJA. A truly general stationary, axially symmetric metric is described by five arbitrary functions of $(r,\theta)$, not the more limited set accessible via the NJA. To establish a more general result, we now adopt the general formalism of Ref.~\cite{Rezzolla2016}:
\begin{align}
	\begin{split}
		\rd s^2 ={}&- \frac{f(r,\theta) - \omega^2(r,\theta) \sin^2\theta}{\kappa^2(r,\theta)} \rd t^2\\
		& - 2 r\omega(r,\theta)\sin^2\theta \rd t \rd\phi  + r^2 \kappa^2(r,\theta) \sin^2\theta \rd \phi^2 \\ 
		& + \sigma(r,\theta) \left[\frac{\beta^2(r,\theta)}{f(r,\theta)} \rd r^2 + r^2 \rd\theta^2\right],
	\end{split}
	\label{eqn:Rezzolla}
\end{align} 
where $f$, $\beta$, $\sigma$, $\kappa$, and $\omega$ are five dimensionless functions of $(r,\theta)$. To avoid any ``metric issue,'' the functions $f$, $\beta$, $\sigma$, and $\kappa$ are required to be positive definite outside any event horizon, if one exists.

For comparison, in the Kerr geometry, these five functions take the following forms:
\begin{align}
	\begin{split}
		f(r,\theta) &= 1 + \frac{a^2-2 M r}{r^2} -\frac{3 a^2 M^2 \sin^2\theta}{\left(r^2 + a^2 \cos^2 \theta\right)^2},\\
		\beta^2(r,\theta) &= 1-\frac{3 a^2 M^2 r^2 \sin ^2\theta}{\left(a^2+r^2-2 Mr \right) \left(r^2 + a^2 \cos^2 \theta\right)^2},\\
		\kappa^2(r,\theta) &= 1 + \frac{a^2}{r^2} + \frac{2 a^2 M \sin ^2\theta}{r\left(r^2 + a^2 \cos^2\theta\right)},\\
		\sigma(r,\theta) &= 1+ \frac{a^2 \cos^2\theta}{r^2},\\
		\omega(r,\theta) &= \frac{a M}{r^2 + a^2 \cos^2\theta}.
	\end{split}
\end{align}
In the spacetime generated by NJA in Eq.~(\ref{eqn:metric}), they are
\begin{align}
	\begin{split}
		f(r,\theta) &= \frac{\Delta(r)}{r^2\rho^4}\Psi^2(r,\theta),\\
		\beta^2(r,\theta) &= \frac{1}{\rho^4}\Psi^2(r,\theta),\\
		\kappa^2(r,\theta) &= \frac{1}{r^2\rho^4} \left\{\left[K(r) + a^2\right]^2 + a^2 \Delta(r) \sin^2\theta \right\} \Psi(r,\theta),\\
		\sigma(r,\theta) &= \frac{1}{r^2}\Psi(r,\theta),\\
		\omega(r,\theta) &= \frac{a}{r}\frac{K(r)-H(r)F(r)}{\rho^4}\Psi(r,\theta).
	\end{split}
\end{align}

Furthermore, inspired by the structure of the Kerr metric in the massless limit $(M\to0)$,
\begin{align}
	\begin{split}
		x & = \sqrt{r^2 + a^2} \sin\theta \cos\phi, \\
		y & = \sqrt{r^2 + a^2} \sin\theta \sin\phi, \\
		z & = r \cos\theta,
	\end{split}
\end{align}
we consider a slightly generalized scalar field ansatz to account for potential rotational asymmetries
\begin{align}
	\begin{split}
		\chi^1 &= \eta h(r) \zeta(r) \sin\theta \cos \phi,\\
		\chi^2 &= \eta h(r) \zeta(r) \sin\theta \sin \phi, \\
		\chi^3 &= \eta h(r)  \cos\theta,
	\end{split}
	\label{eqn:h-zeta}
\end{align}
where $\zeta(r)$ is an additional profile function to be determined. As before, this ansatz leads to two distinct equations of motion (EoMs): one from the degenerate $\chi^{1,2}$ components  
\begin{align}
	\begin{split}
		&  \frac{h\zeta}{r^2\sigma \sin^2\theta} \left[ \frac{\sigma}{\kappa^2} \left(\frac{\omega^2\sin^2\theta}{f} - 1\right)  +\frac{\partial_\theta \left(\beta\sin\theta\cos\theta \right)}{\beta}
		\right] \\
		& + \frac{1}{r^2  \beta \sigma} \partial_r \left[\frac{r^2f}{\beta} \left(h\zeta\right)'\right] \\
		={}& \lambda \eta^2 h \zeta \left[ \left( \cos^2\theta + \zeta^2 \sin^2\theta \right) h^2 -1\right],
	\end{split}
	\label{eqn:EoM12}
\end{align} 
 and another from the $\chi^3$ component,
\begin{align}
	\begin{split}
		&-  \frac{h}{r^2\sigma} \left( 2 +\tan\theta\frac{\partial_\theta \beta}{\beta}
		\right) + \frac{1}{r^2\beta\sigma} \partial_r \left(\frac{r^2f}{\beta} h'\right) \\
		={}&\lambda \eta^2 h \left[ \left( \cos^2\theta + \zeta^2 \sin^2\theta \right) h^2 -1\right].
	\end{split}
	\label{eqn:EoM3}
\end{align} 

The full Einstein tensor for the metric (\ref{eqn:Rezzolla}) is exceedingly complex, see Appendix \ref{sec:Gmunu}. Here, we focus on the energy-momentum tensor, $T_{\mu\nu}$, derived from the ansatz (\ref{eqn:h-zeta}). Its structure dictates the required form of $G_{\mu\nu}$. The non-vanishing off-diagonal components are
\begin{align}
	\begin{split}
		T^t_\phi &= -\eta^2 h^2\zeta^2 \frac{\omega^2 \sin^2\theta}{rf},\\
		T^r_\theta &= \eta^2 \frac{hf \sin\theta\cos\theta}{\beta^2\sigma} \left[\zeta (h\zeta)' - h'\right].		
	\end{split}
\end{align}
The diagonal components are lengthy; for brevity, we present $T^t_t$ explicitly: 
\begin{align}
	\begin{split}
		T^t_t ={}& -\frac{\eta^2}{2} \left\{\frac{h^2\zeta^2}{r^2\kappa^2} \left(1 - \frac{\omega^2 \sin^2\theta}{f}\right) \right.\\
		& + \frac{h^2}{r^2\sigma} \left(\sin^2\theta + \zeta^2 \cos^2\theta \right) \\
		& \left. + \frac{f}{\beta^2\sigma} \left[\left(h'\right)^2 \cos^2\theta + \left(h\zeta\right)'^2 \sin^2\theta\right]\right\} \\
		& - \frac{\lambda \eta^4}{4} \left[ \left(\cos^2\theta + \zeta^2 \sin^2\theta \right)h^2 - 1\right]^2,
	\end{split}
\end{align}
and the others as differences:
\begin{align}
	\begin{split}
		T^r_r &= T^t_t + \eta^2 \frac{f}{\beta^2\sigma} \left[\left(h'\right)^2 \cos^2\theta + \left(h\zeta\right)'^2 \sin^2\theta\right] , \\
		T^\theta_\theta &= T^t_t + \eta^2 \frac{h^2}{r^2\sigma} \left(\sin^2\theta + \zeta^2 \cos^2\theta \right), \\
		T^\phi_\phi &= T^t_t + \eta^2 \frac{h^2\zeta^2}{r^2\kappa^2} \left(1 - \frac{\omega^2 \sin^2\theta}{f}\right).
	\end{split}
\end{align}	
	
Directly solving this full system of eight coupled, non-linear PDEs is intractable. We therefore turn to an asymptotic analysis at large radial distance $r \to \infty$. We expand the seven unknown functions in inverse powers of $r$:
\begin{align}
	\begin{split}
		h(r) = \sum_{n=0}^{\infty} \frac{H_n}{r^n},\quad\,
		& \qquad \zeta(r) = \sum_{n=0}^{\infty} \frac{Z_n}{r^n}, \\
		f(r,\theta) = \sum_{n=0}^{\infty} \frac{F_n(\theta)}{r^n},
		& \qquad	\beta(r,\theta) = \sum_{n=0}^{\infty} \frac{B_n(\theta)}{r^n}, \\
		\kappa(r,\theta) = \sum_{n=0}^{\infty} \frac{K_n(\theta)}{r^n},
		& \qquad		\sigma(r,\theta) = \sum_{n=0}^{\infty} \frac{S_n(\theta)}{r^n}, \\
		\omega(r,\theta) = \sum_{n=0}^{\infty} \frac{O_n(\theta)}{r^n}.
	\end{split}
\end{align}
At leading order $(\mathcal{O}(r^0))$, both EoMs reduce to the algebraic constraint: 
\begin{align}
	 \lambda \eta^2 H_0 \left[H_0^2 \left(\cos^2\theta+Z_0^2 \sin^2\theta\right) - 1 \right] =0.
\end{align}
This equation requires $H_0=0$ or $H_0^2=1$ and $Z_0^2=1$. Note that both EoMs (\ref{eqn:EoM12}) and (\ref{eqn:EoM3}) are invariant under changes in the sign of $\zeta$ or $h$; moreover, all components of the energy-momentum tensor are even functions of $\zeta$ and $h$, and the Einstein tensor does not depend on either $\zeta$ or $h$. Therefore, without loss of generality, we can choose $\zeta(r) > 0$, i.e., $Z_0=+1$. The solution $h(r) = 0$ corresponds to the vacuum configuration without a global monopole. Furthermore, the dynamical equations are invariant under $h \leftrightarrow -h$, implying that the so-called phantom global monopole solution can be obtained simply by reversing the sign of the ordinary monopole solution with $h(r) > 0$. Henceforth, we choose $H_0=+1$ for simplicity. With these choices, the next-to-leading order EoMs reads
\begin{align}
	2 \lambda \eta^2   \left(H_1+Z_1 \sin^2\theta \right) =0.
\end{align}
The only consistent solution is $H_1=0=Z_1.$ 

To proceed further appears daunting. However, the system can be solved by making a single, well-justified assumption leveraging coordinate freedom. We assume the leading-order term of $\beta(r,\theta)$ is independent of $\theta$:
\begin{align}
	B_0(\theta) = B_0.
	\label{eqn:B0}
\end{align} 
This choice is permissible as we can use coordinate transformations to fix one of the metric functions. For instance, the Kerr solution is recovered in this formalism by fixing  $\sigma(r,\theta) = 1 + \frac{a^2}{r^2} \cos^2\theta$ \cite{Rezzolla2016}.  

With this assumption, solving the field equations order by order yields a definitive result. By expanding the off-diagonal equations to $\mathcal{O}\left(r^{-1}\right)$ and the diagonal equations together with EoMs to $\mathcal{O}\left(r^{-2}\right)$, we obtain the correction to the monopole profile:
\begin{align}
	H_2 = - \frac{1}{\lambda \eta^2 K_0^2 r^2}, \qquad Z_2=0,
\end{align}
and a unique leading-order behavior for the metric:
\begin{align}
	\begin{split}
		F_0(\theta) = \left(1-8\pi \eta^2 \right) B_0^2,& \qquad K_0(\theta) = K_0, \\
		S_0(\theta) = K_0^2, & \qquad O_0(\theta) =0.
	\end{split}
\end{align}
This constitutes the central result of our analysis. It demonstrates that the only consistent asymptotic solution requires all leading-order metric functions to be independent of $\theta$ and, crucially, demands that the rotation function vanish ($O_0=0$). The solution is therefore asymptotically spherically symmetric and non-rotating. This pattern persists at all higher orders. The absence of rotation also leads to $\sigma(r,\theta) = \kappa^2(r,\theta)$ to all orders. 

The resulting series solution is a byproduct of our asymptotic analysis. To match the convention in the literature, we choose $K_n=0$ for $n\geq1$ and set the remaining free parameters in the metric as,
\begin{align}
	F_1=-2M, \quad B_0 = 1, \quad K_0=1.
\end{align}
With these choices, the monopole profile becomes
\begin{align}
	\begin{split}
		h(r) ={}& 1 - \frac{1}{\lambda \eta^2 r^2} -\frac{3-16\pi\eta^2}{2 \lambda^2 \eta^4 r^4} 
		%+ \frac{4M}{ \lambda^2 \eta^4 r^5} + \mathcal{O}\left(\frac{1}{r^5}\right), 
		+\cdots ,\quad 
		\zeta(r)=1,	
	\end{split}
\end{align}
and the metric functions are
\begin{align}
	\begin{split}
		f(r,\theta) ={}& 1 - 8\pi \eta^2 - \frac{2M}{r} - \frac{8 \pi}{\lambda r^2} 
		%\\
		%& 
		- \frac{8\pi\left(1-8\pi\eta^2\right)}{3\lambda^2 \eta^2 r^4} +\cdots \\
		%+ \mathcal{O}\left(\frac{1}{r^5}\right)\\
		\beta(r,\theta) ={}& 1 - \frac{4 \pi }{\lambda^2 \eta^2 r^4} +  %\mathcal{O}\left(\frac{1}{r^5}\right)
		\cdots\\
		\kappa(r,\theta) ={}& 1, \quad \sigma(r,\theta) = 1, \quad 
		\omega(r,\theta) ={} 0.
	\end{split}
\end{align}
The analysis confirms that no rotating generalization exists, even within this more general framework. The leading term in $f{(r,\theta)}$ corresponds to the well-known solid angle deficit, while its $\mathcal{O}(r^{-1})$ term can be interpreted as the mass of a central object (e.g., a black hole). The physical significance of the higher-order corrections will be studied in a follow-up paper.

\section{Conclusion}

In this work, we have analyzed why stationary, rotating Barriola-Vilenkin global monopoles do not exist in Einstein's general relativity. Our analysis proceeded in two stages.

First, we investigated the general class of rotating metrics generated by the Newman-Janis algorithm. By rigorously analyzing the coupled Einstein-scalar field equations, we demonstrated a fundamental inconsistency in the scalar field's equations of motion, proving that no rotating solution can be constructed within this framework.

Second, to establish a more powerful and general result, we considered the most general stationary, axially symmetric spacetime. Through detailed asymptotic analysis at large distances, we proved that the only solution that is both regular and consistent with the field equations is the static, spherically symmetric BV monopole. This analysis showed, order by order, that any deviation from spherical symmetry—including rotation—is forbidden.

These two independent lines of reasoning provide compelling evidence that the only stationary, axisymmetric BV global monopole solution is the spherically symmetric one. This suggests that non-trivial rotating BV global monopoles cannot be realized in general relativity.
\begin{acknowledgments}
	M.Y.L.~is supported by the National Natural Science Foundation of China with Grant No.~12305064 and Jiangxi Provincial Natural Science Foundation with Grant No.~20224BAB211020. Q.-h.W.~thanks Prof.~Gong Jiangbin for reading the manuscript and providing valuable feedback.
\end{acknowledgments}

\appendix
\section{Einstein Tensor for the Metric in Eq.~(\ref{eqn:Rezzolla})}
\label{sec:Gmunu}

In this appendix, we list the explicit expression for Einstein tensor $G_{\mu\nu}$ in the generic static, rotational symmetric spacetime defined in Eq.~(\ref{eqn:Rezzolla}): 
\begin{widetext}
\begin{align}
	\begin{split}
		G_{tt} ={}& - \frac{f^2}{r^2\beta^2\kappa^2\sigma} \left\{ \left(1 + r\partial_r\ln\kappa\right)^2 + r \left(\partial_r\ln\frac{f}{\beta^2} \right) \left[1 + \frac{r}{4}\partial_r\ln \left(\kappa^2\sigma \right) \right]
		+ \frac{1}{2} \left(r\partial_r + r^2\partial_{rr}\right) \ln(\kappa^2 \sigma) \right\},\\
		& + \frac{f}{r^2\kappa^2\sigma} \left[1 - \left(\partial_\theta\ln\kappa\right)^2- \frac{1}{4}\left(\partial_\theta\ln\frac{f}{\beta^2}\right) \left(\partial_\theta\ln\frac{f}{\beta^2\kappa^2\sigma}\right) + \frac{1}{2} \partial_{\theta\theta} \ln\frac{f}{\beta^2 \kappa^2\sigma}\right] \\
		&+ \frac{\cot\theta}{r^2\kappa^2\sigma}\left(\frac{f}{2} \partial_\theta \ln\frac{f}{\beta^2\kappa^4} -3\omega^2 \sin^2\theta \partial_\theta\ln\frac{\omega}{\kappa^2}\right) 
		+\frac{f\omega^2\sin^2\theta}{r^2\beta^2\kappa^2
		\sigma} \left[\frac{3}{2}\left( 1 + 2 r \partial_r \ln \kappa\right) + \frac{1}{4} \left(1 - r \partial_r\ln\frac{f}{\kappa^2}\right)^2 
		\right. \\
		& \left.  - \frac{r}{2}   \left(\partial_r\ln\frac{\omega}{\kappa^2}\right) \left(1 + \frac{r}{2} \partial_r\ln \frac{\kappa^6\omega^5}{f^2} \right)  + r \left(\partial_r\ln\frac{f}{\beta^2}\right) \left(1- \frac{r}{4} \partial_r\ln\frac{\omega^2}{f\kappa^2\sigma} \right)- \frac{1}{2} \left( r\partial_r + r^2  \partial_{rr}\right) \ln\frac{\omega^2}{f\kappa^2\sigma}\right] \\
		& + \frac{\omega^2\sin^2\theta}{4 r^2\kappa^2\sigma} \left[ \left(\partial_\theta \ln\frac{f}{\kappa^2}\right)^2  -  \left(\partial_\theta\ln\frac{\omega}{\kappa^2}\right) \left(\partial_\theta\ln\frac{\kappa^6\omega^5}{f^2}\right) + \left(\partial_\theta\ln\frac{f}{\beta^2}\right) \left(\partial_\theta\ln\frac{\omega^2}{\beta^2\kappa^2\sigma}\right)
		+ 2 \partial_{\theta\theta} \ln\frac{\omega^2}{\beta^2\kappa^2\sigma}
		\right]\\
		& - \frac{3\omega^4\sin^4\theta}{4r^2 \kappa^2 \sigma} \left[ \frac{1}{\beta} \left( 1 - r \partial_r\ln\frac{\omega}{\kappa^2}\right)^2 + \frac{1}{f} \left(\partial_\theta\ln\frac{\omega}{\kappa^2}\right)^2
		\right],
	\end{split}\\
%\end{align}
%
%\begin{align}
	\begin{split}
		G_{t\phi} ={}& \frac{3\omega\sin\theta\cos\theta}{2r\sigma} \partial_\theta\ln\frac{\omega}{\kappa^2} - \frac{f\omega\sin^2\theta}{r\beta^2\sigma} \left[ \frac{3r}{2} \partial_r\ln\frac{f}{\beta} + \left( 1 -\frac{r}{2} \partial_r\ln \frac{f}{\kappa^2} \right)^2 - \frac{r}{2} \left(\partial_r\ln \frac{\omega}{\kappa^2} \right) \left( 1 + \frac{r}{2} \partial_r\ln \frac{\kappa^4\omega^2}{f} \right) \right.\\
		& \left.
		- \frac{r^2}{4} \left(\partial_r\ln \frac{f}{\beta^2} \right) \left(\partial_r \ln \frac{\omega}{f\sigma}\right) - \frac{1}{2} \left(r\partial_r +r^2 \partial_{rr}\right) \ln \frac{\omega}{f\sigma}  \right] \\
		&- \frac{\omega\sin^2\theta}{4 r\sigma} \left[ \left(\partial_\theta \ln \frac{f}{\kappa^2}\right)^2 - \left(\partial_\theta \ln \frac{\omega}{\kappa^2} \right) \left(\partial_\theta \ln \frac{\kappa^4\omega^2}{f} \right) + \left(\partial_\theta \ln\frac{f}{\beta^2} \right) \left(\partial_\theta \ln \frac{\omega}{\beta^2\sigma}\right) - 2 \left(\partial_{\theta\theta} \ln\frac{\omega}{\beta^2\sigma} \right) \right] \\
		& +\frac{3\omega^3\sin^4\theta}{4r\sigma} \left[\frac{1}{\beta^2} \left(1 - r\partial_r \ln\frac{\omega}{\kappa^2}\right)^2 + \frac{1}{f} \left(\partial_\theta\ln\frac{\omega}{\kappa^2}\right)^2\right],
	\end{split}\\
%\end{align}
%
%\begin{align}
\begin{split}
	G_{rr} ={}& \frac{1}{r^2} \left\{ 1 + r \partial_r\ln \frac{f}{\kappa^2} -r^2 \left(\partial_r\ln \kappa \right)^2 + \frac{r}{2} \left[\partial_r \ln \left(\kappa^2\sigma \right)\right]\left(1 + \frac{r}{2} \partial_r\ln f \right) \right\}
	\\
	& - \frac{\beta^2}{r^2f} \left[1 - \frac{1}{4} \left(\partial_\theta \ln \frac{f}{\kappa^2}\right)^2 
	- \frac{1}{4} \left(\partial_\theta\ln f\right) \left(\partial_\theta\ln \frac{\kappa^2}{\sigma}\right)  - \frac{1}{2} \partial_{\theta\theta}\ln f \right] +\frac{\beta^2 \cot\theta}{2r^2f} \partial_\theta\ln\frac{f\kappa^2}{\sigma} \\
	& + \frac{\beta^2\omega^2\sin^2\theta}{4r^2f} \left[\frac{1}{\beta^2} \left(1-r\partial_r \ln\frac{\omega}{\kappa^2}\right)^2 - \frac{1}{f} \left(\partial_\theta\ln\frac{\omega}{\kappa^2}\right)^2\right],
\end{split}\\
%\end{align}
%
%\begin{align}
	\begin{split}
		G_{r\theta} ={}& - \frac{1}{2r} \partial_\theta \ln\frac{\kappa^2}{\beta^2\sigma}  - \frac{1}{4} \left(\partial_r\ln f \right) \left(\partial_\theta\ln\frac{f}{\beta^2\kappa^2\sigma} \right) - \frac{1}{4}  \left(\partial_r\ln\frac{f}{\kappa^2\sigma} \right) \left(\partial_\theta\ln f \right) -2 \left(\partial_r\ln \kappa \right)\left(\partial_\theta\ln \kappa \right) \\
		& - \frac{1}{2} \partial_{r\theta}\ln f -\frac{\cot\theta}{2} \partial_r\ln\frac{\kappa^2}{\sigma} - \frac{\omega^2\sin^2\theta}{2rf} \left(1 - r\partial_r\ln\frac{\omega}{\kappa^2}\right) \left(\partial_\theta \ln\frac{\omega}{\kappa^2}\right),
	\end{split}\\
%\end{align}
%
%\begin{align}
	\begin{split}
		G_{\theta\theta} ={}&  \frac{f}{\beta^2} \left[ \frac{r^2}{4}\left(\partial_r \ln\frac{f}{\kappa^2}\right)^2 + \frac{r}{2} \left(  \partial_r \ln \frac{f\kappa^2}{\beta^2\sigma} \right) \left(1 + \frac{r}{2} \partial_r \ln f \right)  + \frac{1}{2}\left(r\partial_r + r^2\partial_{rr}\right) \ln f \right]  - \frac{1}{4}\left(\partial_\theta \ln\frac{f}{\kappa^2}\right)^2
		\\
		& - \frac{1}{4} \left[\partial_\theta \ln \left(f\sin^2\theta \right)\right] \left(\partial_\theta \ln \frac{\kappa^2}{\beta^2\sigma}\right)-\frac{\omega^2 \sin^2\theta}{4} \left[\frac{1}{\beta^2} \left(1 - r \partial_r \ln\frac{\omega}{\kappa^2}\right)^2 -\frac{1}{f} \left(\partial_\theta \ln\frac{\omega}{\kappa^2}\right)^2\right],
	\end{split}\\
%\end{align}
%
%\begin{align}
	\begin{split}
		G_{\phi\phi} ={}& \frac{f \kappa^2 \sin^2\theta}{\beta^2\sigma} \left[ \frac{r^2}{4} \left(\partial_r\ln\frac{f}{\kappa^2}\right)^2 + \frac{r}{2} \left(\partial_r\ln\frac{f}{\beta^2}\right)\left( 1 + \frac{r}{2} \partial_r\ln\frac{f\sigma}{\kappa^2}\right) + \frac{1}{2}\left( r\partial_r + r^2
		 \partial_{rr}\right) \ln\frac{f\sigma}{\kappa^2}\right]\\
		& + \frac{\kappa^2\sin^2\theta}{\sigma} \left[ \left(\partial_\theta\ln\frac{\beta}{\kappa}\right)^2 +\frac{1}{4} \left(\partial_\theta\ln\frac{f}{\beta^2}\right) \left(\partial_\theta\ln\frac{f}{\kappa^2\sigma}\right) -\frac{1}{2} \partial_{\theta\theta} \ln\frac{\kappa^2}{\beta^2\sigma}\right] \\
		& - \frac{3 \kappa^2 \omega^2 \sin^4\theta}{4\sigma} \left[\frac{1}{\beta^2} \left(1-r\partial_r\ln\frac{\omega}{\kappa^2}\right)^2 + \frac{1}{f} \left(\partial_\theta\ln\frac{\omega}{\kappa^2}\right)^2 \right]  .
	\end{split}
\end{align}
All other components of $G_{\mu\nu}$ are zero.
\end{widetext}


\begin{references}

%Global monopoles
\bibitem{Kibble1976}
	T.~W.~B.~Kibble, \textit{Topology of cosmic domains and strings}, J.~Phys.~A \textbf{9}, 1397 (1976).
\bibitem{Vilenkin1985}
	A.~Vilenkin, \textit{Cosmic strings and domain walls}, Phys.~Rep.~\textbf{121}, 263 (1985).
\bibitem{Vilenkin1989}
	M.~Barriola and A.~Vilenkin, \textit{Gravitational field of a global monopole}, Phys.~Rev.~Lett.~\textbf{63}, 341 (1989).
\bibitem{Vilenkin1994}
	A.~Vilenkin and S.~Shelard, \emph{Cosmic strings and Other Topological Defects.} (Cambridge University Press, Cambridge 1994).

%GM Reviews
\bibitem{Kibble1995}
M.~B.~Hindmarsh, T.~W.~B.~Kibble, \textit{Cosmic strings}, Rep.~Prog.~Phys.~\textbf{58}, 477 (1995).

\bibitem{Perlick2004}
	V.~Perlick, \textit{Gravitational Lensing from a Spacetime Perspective}, Living Rev.~Relativity \textbf{7}, 9 (2004).

\bibitem{Tsujikawa2006}
	E.~J.~Copeland, M.~Sami, and S.~Tsujikawa, \textit{Dynamics of Dark Energy}, Int.~J.~Mod.~Phys.~D \textbf{15}, 1753 (2006).

\bibitem{Radu2015}
	C.~A.~R.~Herdeiro and E.~Radu, \textit{Asymptotically flat black holes with scalar hair: A review}, Int.~J.~Mod.~Phys.~D \textbf{24}, 1542014 (2015). 

%\bibitem{EHT2023}
%	S.~Vagnozzi \textit{et al}, \textit{Horizon-scale tests of gravity theories and fundamental physics from the Event Horizon Telescope image of Sagittarius A*}, Class.~Quantum Grav.~\textbf{40}, 165007 (2023).

%Rotating GM

\bibitem{Bezerra2001}
R.~M.~Teixeira Fihlo and V.~B.~Bezerra, \textit{Gravitational field of a rotating global monopole}, Phys.~Rev.~D \textbf{64}, 084009 (2001).

\bibitem{Bezerra2012}
J.~P.~M.~Gra\c{c}a and V.~B.~Bezerra, \textit{Gravitational Field of a Rotating Global Monopole in $f(R)$ Theory}, Mod.~Phys.~Lett.~A \textbf{27}, 1250178 (2012).

\bibitem{Jing2013}
S.~Chen and J.~Jing, \textit{Gravitational field of a slowly rotating black hole with
	a phantom global monopole}, Class.~Quantum Grav.~\textbf{30} 175012 (2013).

\bibitem{Jusufi2017}
K.~Jusufi, M.~C.~Werner, A.~Banerjee, and A.~\"Ovg\"un, \textit{Light deflection by a rotating global monopole spacetime}, Phys.~Rev.~D, \textbf{95}, 104012 (2017).

\bibitem{Jusufi2020}
Su.~Haroon, K.~Jusufi, and M.~Jamil, \textit{Shadow Images of a Rotating Dyonic Black Hole with
	a Global Monopole Surrounded by Perfect Fluid}, Universe, \textbf{6}, 23 (2020).

\bibitem{Asada2019}
T.~Ono, A.~Ishihara, and H.~Asada, \textit{Deflection angle of light for an observer and source at finite distance from a rotating global monopole}, Phys.~Rev.~D \textbf{99}, 124030 (2019).

\bibitem{Fathi2025}
M.~Fathi, \textit{Shadow Analysis of an Approximate Rotating Black Hole	Solution with Weakly Coupled Global Monopole Charge}, Universe \textbf{11}, 111 (2025).

%debate
\bibitem{Secuk2020}
M.~Haluk Se\c{c}uk, \textit{Comment on ``Gravitational field of a rotating global monopole,''} Phys.~Rev.~D \textbf{102}, 068501 (2020).

\bibitem{Bezerra2020}
V.~B.~Bezerra, J.~M.~Toledo, J.~P.~Morais Gra\c{c}a, D.~Barbosa, and R.~M.~Teixeira Filho, \textit{Reply to ``Comment on `Gravitational field of a rotating global monopole,' ''} Phys.~Rev.~D \textbf{102}, 068502 (2020).


%NJA	
\bibitem{Newman1965a}
 	E.~T.~Newman and A.I.~Janis, \textit{Note on the Kerr Spinning-Particle Metric}, J.~Math.~Phys.~\textbf{6}, 915 (1965).
\bibitem{Newman1965b}
	E.~T.~Newman, E.~Couch, K.~Chinnapared, A.~Exton, A.~Prakash, R.~Torrence, \textit{Metric of a Rotating, Charged Mass}, J.~Math.~Phys.~\textbf{6}, 918 (1965).
\bibitem{Szekeres2000}
	S.~P.~Drake and P.~Szekeres, \textit{Uniqueness of the Newman–Janis Algorithm in 		Generating the Kerr–Newman Metric}, Gen.~Rel.~Grav.~\textbf{32}, 445 (2000). 

%Review on NJA
\bibitem{Erbin2017}
	H.~Erbin, \textit{Janis–Newman Algorithm: Generating Rotating and NUT Charged Black Holes}, Universe \textbf{3}, 19 (2017).

\bibitem{AA2014a}
	M.~Azreg-A\"{i}nou, \textit{Regular and conformal regular cores for static and rotating solutions}, Phys.~Lett.~B \textbf{730} 95, (2014).
\bibitem{AA2014b}
	M.~Azreg-A\"{i}nou, \textit{From static to rotating to conformal static solutions: rotating
		imperfect fluid wormholes with(out) electric or magnetic field}, Eur.~Phys.~J.~C \textbf{74}, 2865 (2014).
\bibitem{AA2014c}
	M.~Azreg-A\"{i}nou, \textit{Generating rotating regular black hole solutions without complexification}, Phys.~Rev.~D \textbf{90}, 064041 (2014).


%General axisymmetric metric

\bibitem{Rezzolla2016} 
	R.~Konoplya, L.~Rezzolla, and A.~Zhidenko, \textit{General parametrization of axisymmetric black holes in metric theories of gravity}, Phys.~Rev.~D \textbf{93}, 064015 (2016).


%\bibitem{Gangopadhyay2008}
%	S.~Gangopadhyay, \textit{Hawking radiation from a Reissner-Nordstr\"om black hole with a global monopole via covariant anomalies and effective action}

%\bibitem{18}R.Nobili, The Conformal Universe I,arXiv:1201.2314


\end{references}
\end{document}